# Participation of Stakeholder in the Design of a Conception Application of Augmentative and Alternative Communication

Flavien Clastres-Babou[1], Frédéric Vella[1], Nadine Vigouroux[1], Philippe Truillet[1], Nadine Vigouroux[1], Charline Calmels[2], Caroline Mercadier[2], Karine Gigaud[2], Margot Issanchou[2], Kristina Gourinovitch[2], Anne Garaix[2]

[1] IRIT Laboratory, Paul Sabatier University, Toulouse, France
[2] OPTEO Foundation, Onet-le-Château, France
Nadine.Vigouroux@irit.fr

**Abstract.** The objective of this paper is to describe the implication of an interdisciplinary team involved during a user-centered design methodology to design the platform (WebSoKeyTo) that meets the needs of therapists to design augmentative and alternative communication (AAC) aids for disabled users. We describe the processes of the design process and the role of the various actors (therapists and human computer researchers) in the various phases of the process. Finally, we analyze a satisfaction scale of the therapists on their participation in the codesign process. This study demonstrates the interest in extending the design actors to other therapists and caregivers (professional and family) in the daily life of people with disabilities.

## 1 Context

Assistive technologies for communication and home automation allow people with disabilities to be autonomous and better social participation. However, many of these assistive technologies are abandoned [1] because they do not sufficiently take into account the expression of the needs of these people. In order for these technologies to meet the needs, it is important to involve, in user-centered design, occupational therapists and psychologists who can complement or express the needs of people with disabilities as part of their ecosystem [2]. Moreover, their expertise allows them to evaluate the abilities of the disabled person to better adapt the assistive technologies to their needs.

In previous work [3], experts in human-computer interaction developed AAC with the SoKeyTo platform [4] based on the needs expressed by therapists. It consists of a page and button editor for the use of AAC designers.. It consists of a page and button editor for the use of AAC designers. The editor allows defining the morphology and the contents of the buttons (text, picture and sound), the visual and sound feedback and the type of associated function (simple communication function, call of an application, sending of messages according to protocols to restore an oral message with a text-to-speech system) and the structuring of buttons in page. The SoKeyTo platform editor allows the customization of the AAC interface and the connection of several input interaction modes (pointing device, eye tracker, joystick, speech recognition, on/off switch). SoKeyTo also allows various control modes to be configured (pointing, time delay click, scanning system) according to the abilities of disabled people.

Calmels et al. [4] reported the limits of the design of AACs by human-computer interaction engineers (difficulty in understanding the needs given by the therapists; longer design time and availability of AAC due to numerous exchanges between designers and therapists; impossibility to test AAC adaptations online). This study also demonstrated the crucial role of occupational therapists in the learning phase and the adaptation of AAC to abilities and behavior of the disabled person. Indeed, occupational therapists and psychologists have knowledge and know how to do for adapting and personalizing AAC during the occupational therapy sessions.

Numerous studies have addressed the issue of end-user participation in the codesign of applications. Gibson et al. [5] give recommendations to overcome barriers on how to better support people with disabilities to engage in codesign. Dijks et al. [6] have proposed participation methods that empower people with impairment to actively take part in the design process. Ambe et al. [7] presents codesign fiction as an approach to engaging users in imagining, envisioning and speculating not just on future technology but future life through co-created fictional works. These studies show that different methods of codesign are invented to overcome barriers in order to facilitate the participation, the expression of needs and the proposal of solutions by end users.

Participatory design with therapists without experience of user-centered design method poses challenges for researchers and designer due to the differences in their mutual experiences and knowledge. In this paper, we firstly highlighted the methodological approach adopted for the codesign of WebSoKeyTo, taking into account the experiences of using AAC produced by SoKeyTo. Then, we report the satisfaction questionnaire about the participation of the therapist in the design process of the WebSoKeyTo platform.

## 2   Codesign of the WebSoKeyTo Platform Using a User-Centered Design Method

Figure 1 describes the different phases of the user-centred design methodology [8] of the WebSoKeyTo platform for designing alternative augmentative communication aids. The team consists of six therapists (4 psychologists, 2 occupational therapists) and 5 human-computer interaction researchers (3 senior researchers and 2 students). The six therapists had never participated in a focused design method before. One of the therapists has a dual background (psychologist and computer specialist). The following section describes the phases of the implementation of the method. A satisfaction questionnaire on the participation of the therapists ends this design cycle.

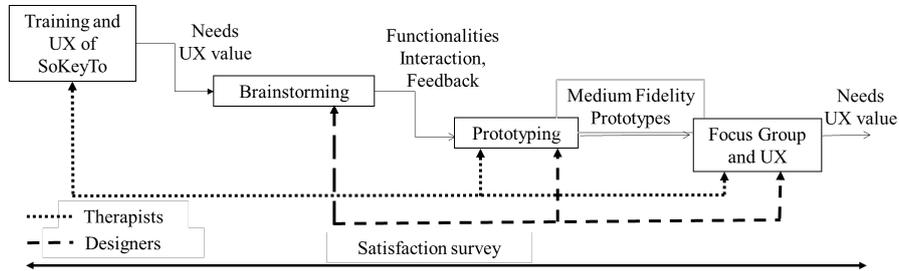

**Fig. 1.** Different phases of the design of the WebSoKeyTo platform.

### 2.1 Training and Evaluation of SoKeyTo

We first trained the six therapists to use the SoKeyTo [3] platform by demonstrating all the features in a practical way as reported above. Five users had never used another AAC design tool before and one is expert in the use of SoKeyTo. These six therapists were invited to use the SoKeyTo platform for two months: firstly, a scenario imposed by the SoKeyTo platform designers for one month, and then a free scenario for the design of an AAC for a disabled person. These therapists could benefit from the help of the SoKeyTo designers in case of bugs or difficulties of use. At the end of this trial phase of the SoKeyTo platform, we proceeded to the evaluation of the usability of this platform by means of the USE (Usefulness, Satisfaction, and Ease of use) questionnaire [9] on a Likert scale from 1 (strong disagree) to 7 (strong agree).

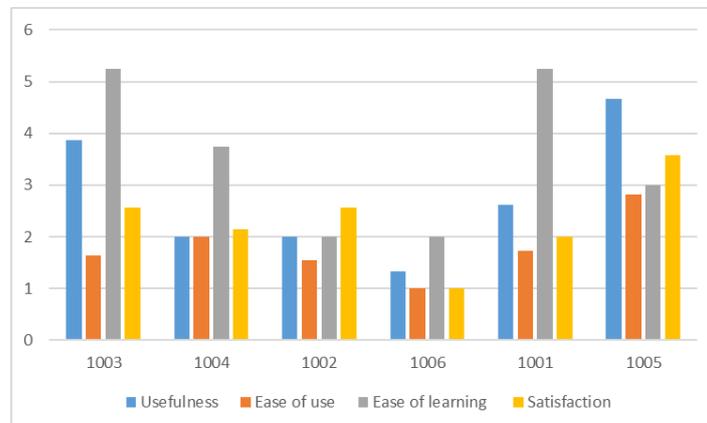

**Fig. 2.** Score of the USE questionnaire for each therapist.

Figure 2 shows the results of the five USE indicators for each of the therapists. We note that professional 1006 (psychologist and former computer scientist) rated the SoKeyTo platform very negatively. However, two other therapists mentioned its ease of learning and three others its ease of use.

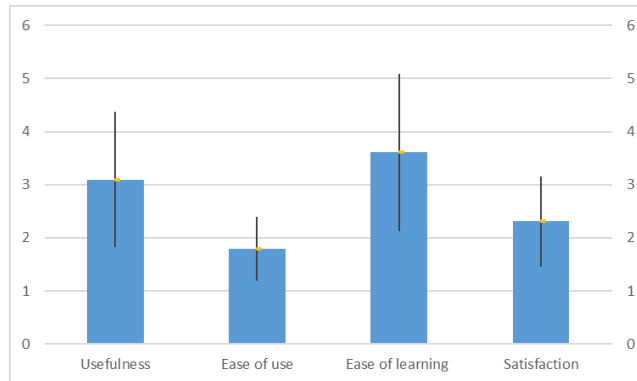

**Fig. 3.** Mean and standard deviation of USE criteria.

Figure 3 shows a very poor score for the criterion "Ease of Use" (mean :1.8 ; standard deviation (SD) :0.6) and for the criterion "Satisfaction" (mean :2,3; SD :0,85. The criterion "Ease of learning" (mean :3,6 ; SD 1.4) and "Usefulness" (mean: 3; SD: 1,3) are just average. Difficulty of understanding (concept too computer-oriented and not professional) of the functionalities, SoKeyTo oriented towards computer use, lack of essential functionalities (undo, overview of the AAC structure, pictogram editor, etc.), bugs, poor ergonomics, aesthetics to be reviewed are negative points reported by the therapists.

The therapists reported positive such as functions to run applications or web links from the buttons, customization of the AAC interface, possible interaction with different switches, high creative potential to create an AAC.

At the end of the SoKeyTo use phase, the six therapists participated in a focus group to express their feedback on the use of SoKeyTo and their needs. Therapists reported mainly the lack of ergonomics of the SoKeyTo editor's interface and lack of functionality. They also expressed additional needs: a Web version of SoKeyTo (WebSoKeyTo), the possibility to share their resources (pictures, pages, AAC interface, ...), a user interface of page and button editor more user-friendly including menus, functionalities, and buttons more accessible for therapists. The therapists have ranked their needs after a consensus between them. Then, the needs were discussed with the whole codesign team. Therapist 1006 played an important role in clarifying the needs in terms of software functionality.

## 2.2 Brainstorming

From the user feedback obtained through the questionnaire and the feedback on SoKeyTo, the design team used this base to design a solution closest to the therapists' needs. The design team set up two brainstormings through several low-fidelity mock-ups. The purpose was to propose a new friendly user interface and a redesign of functionalities available on SoKeyTo more accessible to therapists.

We performed two different ones. The first one concerns the specification of AAC buttons (morphological and functional characteristics) and the specification of the scanning strategy (parameters of scanning strategy) of AAC.

The second one is about the navigation of pages. A page is a set of navigation and communication buttons that can be grouped into categories. This idea came from therapists' activities with communication books used by disabled persons to communicate within their human environment.

### 2.3 Medium Fidelity Prototype

We set up an alternative sequence of prototype design and focus group for the three prototypes (specification of buttons, of scanning system and navigation of pages, see Figure 4).

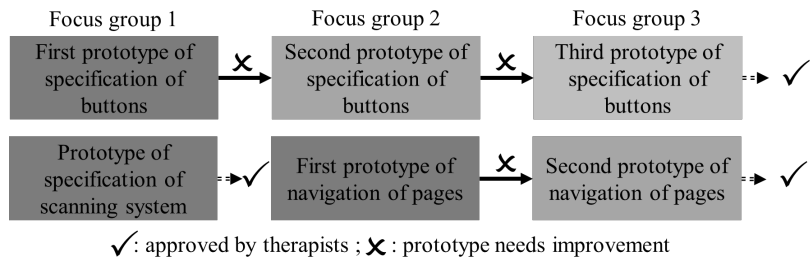

Fig. 4. Focus group and prototyping cycle.

The design team set up focus groups to collect feedback from the therapists on the prototype and new functionalities that allow them to carry out their activities. Due to COVID pandemic, all focus group were performed by videoconferencing. The design team show prototypes using case studies. Then, design team and therapists shared their point of view about prototypes. The both skills of the focus group proposed some adjustments or improvements. At the end of the focus group, if there was no agreement to validate the prototype, it was modified and presented again a week later until consensus.

The first iteration was to present the prototypes on specification of buttons and specification of scanning system. The feedback was very positive, but the therapists wished more concrete examples to have a better overview. The scanning strategy prototype was validated during the first focus group, as it corresponds to the expectations of the users due to its affordance (choice of strategy parameters through visual representations).

The second session was about specification of buttons prototype and navigation of pages. Therapists highlighted their wish to group some functionalities to have an easier control on WebSoKeyTo application for the specification of button. For the navigation of pages prototype, therapists asked to have a more developed case studies to see the envisaged result for a large number of pages. The CCA described in [4] consists of 53 pages. They also suggested modifications concerning the use of a color code to differentiate the category of pages from each other, The last iteration allowed to validate the two prototypes after some minor modifications following the therapists' remarks.

The therapists are testing the prototype of the button specification. They greatly appreciated being able to use it and adjust their needs in terms of feedback and interaction on the buttons. The development of scanning and page navigation strategies is still in progress.

## 3       Study of Therapists' Satisfaction with Their Participation

The therapists were heavily involved throughout the design process.Our objective was to evaluate the satisfaction of participation of 5 therapists (3 psychologists and 2 occupational therapists) in the codesign phase of the WebSoKeyTo platform by means of a questionnaire (See Table 1).

**Table 1.** Satisfaction questionnaire on the involvement of the therapists.

| Number | Questions |
| --- | --- |
| Q1 | Do you feel that you are involved in the codesign? |
| Q2 | Do you feel that your design proposals have been taken into account in the medium proposals? |
| Q3 | Do you feel that your design proposals have been taken into account in the V1 WebSoKeyTo platform? |
| Q4 | Do you feel that your professional skills were taken into account in the codesign? |
| Q5 | Were your proposals taken into account quickly? Why or why not? |
| Q6 | Do you think you had difficulties in expressing ideas? |
| Q7 | Do you think you had difficulties in expressing solutions (concrete proposals)? Why or why not? |
| Q8 | Are you satisfied with the way consensus was reached? Why or why not? |

We used a Likert scale with five values (strongly disagree, disagree, neutral, agree, strongly agree). Figure 5 shows that therapists globally appreciated their participation in the design process (Q1 to Q4, Q8). However, the answers to Q6 and Q7 show that two therapists had difficulties due to the lack of practice and the need to share a common language to express needs and solutions.

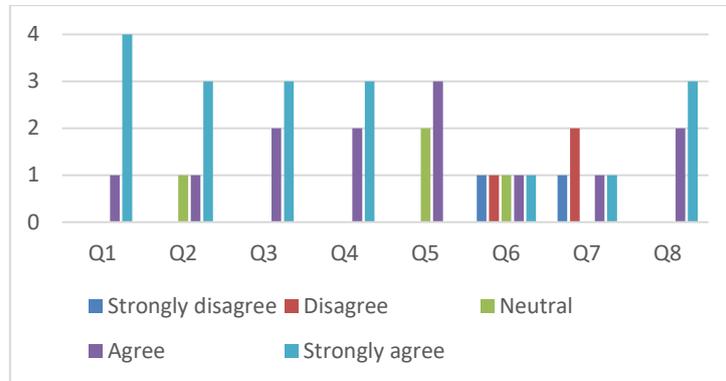

**Fig. 5.** Frequency of responses from the 5 therapists

All of them underline the listening skills of the design team. The proposals discussed between the therapists, the regular and numerous exchanges allowed to find consensus between the needs formulated by the therapist and the functionalities and interaction proposed by the designer team. Sometimes a psychologist, ex-computer scientist, facilitated the understanding of the therapists' proposals.

## 4   Discussion

This questionnaire shows that therapists are satisfied with the way their needs were taken into account and how the design consensus was conducted. However, to the question "which other stakeholder" should be consulted for the design of the WebSoKeyTo platform, the therapists suggest integrating speech therapists specialised in language, psychomotor therapist who could have another point of view. They also suggested that daily carers (specialised educational monitors, educational and social carers, carers, etc.) as well as family members could be involved in the adaptation of AAC, including functionalities/pictograms related to life books. To the question, "how do you improve your involvement in the design of WebSoKeyTo"? Assistance in the testing phases, better explanation of the expectations of the design team are requests expressed by the therapists. They also recommended that a greater immersion of the designers with the professionals could have been beneficial.

We implemented the method of codesigning the WebSoKeyTo platform with two therapists' skills. The discussions show that the people who accompany the disabled person in their daily lives should also be involved in the design process. The therapists suggest that Fablabs could also be involved in the modelling and manufacturing of the control devices. This shows that the therapists are already planning to use the AAC designed by the WebSoKeyTo platform.

## 5      Conclusion

This paper describes a search for an appropriate approach for involving therapists in the design of the WebSoKeyTo platform. None of the therapists had been involved in a design methodology and three of the five researchers were partially aware of the needs of therapists. Firstly, the version used by the engineers underwent training for therapists, followed by two months of use and user experience. This evaluation clearly demonstrated the need to improve the ergonomics, the logic of the AAC design interface and its affordance. This discovery of the SokeyTo platform was essential for the therapists to mature their needs. To do this, we implemented a cycle of codesign tools (focus group, brainstorming, prototyping). This study showed the need to develop a common language and a total immersion of the researchers to understand the therapists' needs. Conversely, the therapists had to take into account the technological and ethical constraints of their request. This study therefore demonstrated the need of a close collaboration and various exchange to find consensus for the WebSoKeyTo design. The medium fidelity prototypes also showed their limitations. Indeed, therapists would have liked to manipulate these prototypes with more representative case studies. The satisfaction questionnaire also shows that the stakeholder needs to be extended to other everyday professionals, especially for the adaptation of the AAC by them. The question of reinventing design methods in the form of method stories [10] or codesign fiction approaches [7] used for the participation of people with disabilities arises for this design context.

**Acknowledgement.** This study is partially funded by the ANS (France), Structures 3.0 tender